\documentclass[12pt]{article}

\usepackage{graphicx}
\usepackage{epsfig}

\usepackage{epsfig, bm}

\RequirePackage{color}
\definecolor{MyDarkGreen}{rgb}{0.02,0.60,0.06}

\title{Critical mass and the dependency of research quality on group size}
\author{ 
 {\it R.~Kenna$^{\,1}$} and {\it B.~Berche$^{\,2}$,} \\~\\
$^1$ Applied Mathematics Research Centre,\\
Coventry University,\\
Coventry, CV1 5FB, England
{}\\~\\
$^2$ Statistical Physics Group,
 Institut Jean Lamour\footnote{Laboratoire associ\'e au CNRS UMR 7198} ,\\
 CNRS -- Nancy Universit\'{e} -- UPVM, B.P. 70239,\\
 F -- 54506 Vand{\oe}uvre l\`es Nancy Cedex, France
{}\\~\\}
\begin{document}
\maketitle

{\Large
  \begin{abstract}
Academic research groups are treated as complex systems and their cooperative behaviour is analysed from a
mathematical and  statistical viewpoint. Contrary to the naive expectation that the 
quality of a research group is simply given by the mean calibre of its individual scientists, we show that 
intra-group interactions play a dominant role. Our model manifests phenomena akin to phase transitions 
which are brought about by these 
interactions, and which facilitate the quantification of the notion of critical mass for research groups.
We present these critical masses for many academic areas. 
A consequence of our analysis is that overall research performance of a given discipline is improved by supporting 
medium-sized groups over large ones, while small groups must strive to achieve critical mass.

  \end{abstract} }
%
  \thispagestyle{empty}
%
%
  \newpage
%
                  \pagenumbering{arabic}

\section{Introduction}

The capacity to assess the relative strengths of research groups is important for research institutes, 
funding councils and governments that must decide on where to focus investment.
In recent years, there have been pressures to concentrate funding on institutions which already have significant resources, 
in terms of finances and staff numbers, due to an expectation that these produce higher quality research \cite{Ha09}. 
On the other hand, advocates of competition argue for a more even spread of 
resources to also support pockets of excellence found in smaller universities.
A central question in this debate \cite{Ha09} is whether there exists a critical mass in research and, if so, what is it? 
Here we show that research quality is indeed correlated with group size and that 
there are, in fact, two significant, related masses, which are discipline dependent. 
The critical mass marks the size below which a group is vulnerable to extinction 
and there is also a higher value at which  the correlation between research quality and group size reduces.
We present a model based on group interactions and determine the critical masses for many academic disciplines.

The notion of relative {\emph{size}} of research groups varies significantly across subject areas. 
Larger research groups are common in experimental disciplines while theorists tend to work 
in smaller teams or even individually. 
Critical mass may be loosely described as the minimum size a research team must attain for it 
to be viable in the longer term.
This is subject dependent. 
For example, life would clearly be very difficult for a single experimental physicist, 
while a pure mathematician may be quite successful working in isolation.
Once critical mass is achieved, a research team has enhanced opportunities for cooperation as well as 
improved access to more resources.  
Compared to pure mathematicians, one expects that a greater critical mass of experimental 
physicists is required to form a viable research team.

Indeed, in recent years there has been a tendency in some countries to concentrate resources into 
larger research groups at the expense of 
smaller ones and to encourage smaller teams to merge, both within and between institutions \cite{Ha09}. 
The question arises as to what extent conglomeration of research
teams influences research quality. 
Larger teams may have an advantage  in terms of environmental dynamism 
(collaboration fosters discussion and vice versa) and 
reduced individual non-research workloads (such as teaching and administration).
Here, we quantify these intuitive notions and the effectiveness of conglomeration of research teams.

This work is primarily based upon the measures of research quality as determined in the UK's Research 
Assessment Exercise (RAE).
Although the data are primarily UK based, this analysis should be of widespread interest \cite{Ki04}. 
For example, the French assessment system, 
which is performed by the {\emph{Agence d'{\'E}valuation de la Recherche et de l'Enseignement Sup{\'e}rieur}} (AERES),
is attempting to move towards a more accurate methodology, similar to the UK system. 
Therefore to check the generality of our analysis, we compare the results of the UK's RAE with those of the French equivalent.

The RAE is an evaluation process undertaken every 5-7 years on behalf of the higher education 
funding bodies in the UK. These bodies use the results of the RAE to allocate funding to universities 
and other higher education institutes for the subsequent years. 
The most recent RAE was carried out in October 2008 and the results were published in March 2009. 
Subject areas were examined to  determine the proportion of research submitted 
which fell into five quality levels. These are defined as:
\begin{itemize}
\item
4*: Quality that is world-leading in terms of originality, significance and rigour 
\item
3*: Quality that is internationally excellent in terms of originality, significance and rigour but which nonetheless falls short of the highest standards of excellence 
\item
2*: Quality that is recognised internationally in terms of originality, significance and rigour 
\item
1*: Quality that is recognised nationally in terms of originality, significance and rigour
\item Unclassified: Quality that falls below the standard of nationally
 recognised work
\end{itemize}
A formula based on the resulting quality profiles
is then used to determine how research funding is distributed. The formula used by the 
Higher Education Funding Council for England  associates each
rank with a weight in such a way that 4* and 3* research  
respectively receive seven and three times the amount of funding 
allocated to 2* research, and 1* and unclassified research attract no funding.
This funding formula therefore represents a measure of quality and can be
used to quantify the quality of a research group.
Based upon this, we  formulate a method to  compare different research groups within and between different disciplines.

In the AERES evaluation system France is geographically divided into four parts, 
one part being evaluated each year. 
The 2008 evaluation, which is the most recent for which data are available, 
is considered more precise than the previous exercise 
and facilitates comparison with the British approach.
However, since only 41 different institutions were  evaluated (and of them, only 10 were traditional universities), 
the amount of data
available for the French system is lower than for the UK equivalent. 
Furthermore, only a global mark is attributed to cumulated research groupings which can include several teams 
with heterogeneous levels.
As a consequence,  we lose the fine-grained analysis at the level of the research teams. 
This is clearly a weak point compared to the British system of evaluation and the 
AERES intends to change it in the near future.
Nonetheless, in order to make the comparison with the British system, we translate 
the AERES grades A+, A, B, C into 4*, 3*, 2*, and 1*. 

From the outset, we mention that there are obvious assumptions underlying our analysis and limits to what 
it can achieve. 
We assume that the RAE scores are reasonably robust and reliable (an assumption which is borne out by our analysis).
We cannot account for collaborations between academia and industry, and we have omitted the engineering 
disciplines from this report as no clear patterns were discernible.
Nor can we account for managerial tactics whereby assessed researchers are relieved from other 
duties such as teaching and administration. 
Factors such as these amount to (sometimes considerable) noise in the systems and it is remarkable 
that, despite them,
reasonably clear and quantitative conclusions can be drawn.

\section{The Relationship Between Quality and Quantity}

It is reasonable to expect that both the size and the quality of a research group are 
affected by a multitude of factors: the calibre of individual researchers, the strength of communication links 
between them, their teaching and administrative loads, the quality of management, 
the extent of interdisciplinarity, the equipment used, whether the work is mainly
experimental, theoretical or computational, the methodologies and traditions of the field,
library facilities, journal access, extramural collaboration,
and even previous successes and prestige factors. 
We will show that of all these and other factors, the dominant driver of research quality is the 
quantity of researchers that an individual is able to communicate with. 
Here, we develop a microscopic model for the relationship between quality and quantity.
It will be seen through rigorous statistical analyses 
that this model captures the relationship well and that quantity of intra-group communication links is 
the dominant driver of group quality. Other factors then contribute to deviations of the qualities
of individual groups from their size-dependent averages or expected values.

We may consider the English funding formula for 2009 as a basis for a 
measurement of the  relative {\emph{strength}} or {\emph{quality}} of 
research groups evaluated at RAE. 
If $p_{n*}$ represents the percentage of  a team's 
research which was rated  $n$*, this gives that team's 
{\emph{quality} as
\begin{equation}
 s = p_{4*} + \frac{3}{7}p_{3*} + \frac{1}{7}p_{2*}
\,.
\label{seven}
\end{equation}
If the team is of size $N$, then the amount of funding 
allocated to it in England is proportional to  the {\emph{absolute strength}} 
$S$ of the team, namely
\begin{equation}
 S= s N  
\,.
\label{Ns}
\end{equation}

Denote the strength of the $i^{\rm{th}}$ member of  group $g$ by ${a_g}_i$. 
This parameter encapsulates not  only the calibre of the individual 
and prestige of the group, but the added strength that individual enjoys due to factors such as extramural collaboration, access to 
resources, teaching loads etc. It does {\emph{not}\/} encapsulate cooperation between individual $i$ and individual $j$, say. 
We address such factors below.
 
A naive view is that the strength $S_g$ of group $g$ is given by the accumulated strengths
of the $N_g$ individuals comprising it, so that
\begin{equation}
 S_g = \sum_{i=1}^{N_g}{{a_g}_i} = N_g \bar{a}_g,
\label{wrong}
\end{equation}
where $ \bar{a}_g$ is the average strength of the individuals in the group.
According to Eq.(\ref{Ns}), the {\emph{quality}\/} of group $g$ is now $s_g = S_g/N_g =  \bar{a}_g$, 
i.e., the group quality coincides with the average strength of group members.

If $\cal{N}$ is the total number of groups in a given discipline, we may now write 
\begin{equation}
 \bar{a} = \frac{1}{\cal{N}}\sum_{g=1}^{\cal{N}} s_g = \frac{1}{\cal{N}}\sum_{g=1}^{\cal{N}}  \bar{a}_g
\end{equation}
to denote the mean group quality. 
According to this naive interpretation, individual group quality scores should be distributed about this
mean value. 
Indeed this is apparent in Fig.~1(a), in which quality measurements for applied mathematics
groups are plotted against the institutes hosting them.
The plot displays a scatter about the mean quality which is also displayed, so that
absolute strength  $S$ is, on average, proportional to group size $N$.
An obvious impression gained from the plot
is that groups whose measured research quality is beneath the mean are underperforming and those whose
quality values lie above the mean are performing well. 
This is the basis on which groups and universities are ranked and the consequences of this interpretation can therefore be enormous.
However, we will show that this impression is dangerous, and, can actually be misleading. 

\begin{figure}[!t]
\begin{center}
\hspace{-1cm}
\includegraphics[width=0.45\columnwidth, angle=0]{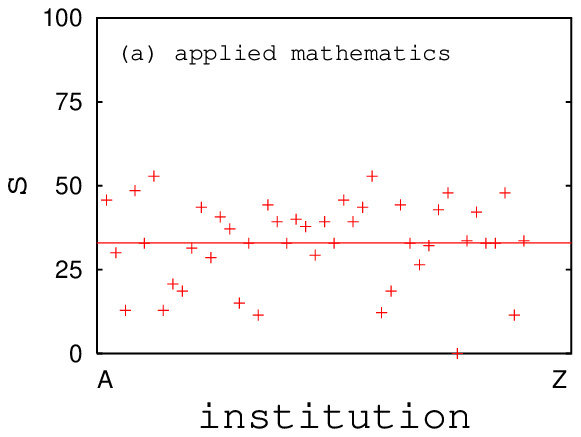}
\includegraphics[width=0.45\columnwidth, angle=0]{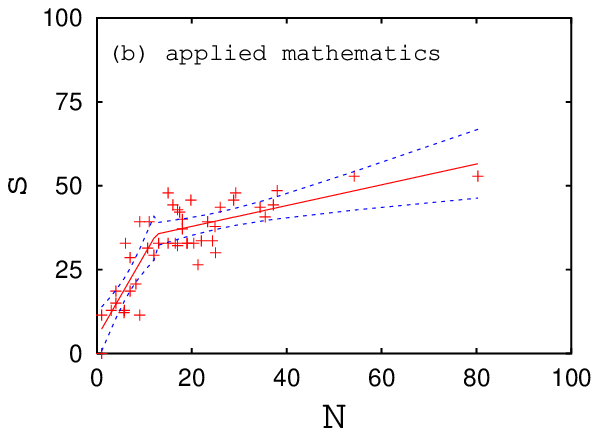}
\caption{Quality measurements for each of the 45 UK applied mathematics groups (a) plotted alphabetically
according to university name
and (b) plotted against group size. In panel (a), the solid line is the mean quality of these 45 groups.
In (b), a correlation between quality and groups size is apparent. The solid line (red online) is a
piecewise linear regression best-fit to the data and the  dashed curves (blue online) represent 95\%  
confidence intervals for this fit.}
\end{center}
\end{figure}

That the impression coming from Fig.~1(a) is  not the full picture is demonstrated in Fig.~1(b), where 
the quality  $s$ for applied mathematics is plotted against group size $N$. 
Clearly the distribution is not entirely random and there is a positive correlation between 
quality $s$ and  the team size $N$: larger teams tend to have higher quality. 
To understand the reason for the correlation between quality and group size, we must 
consider  research groups as {\emph{complex systems}} and take intra-group interactions into account.

The current view within the physics community is that a complex system is
either one whose behaviour crucially depends on the details of the system \cite{Parisi} 
or one comprising many interacting entities of non-physical origin \cite{Yurko}. Here, the second definition
is apt.
In recent years, statistical physicists have turned their attention to the analysis of such systems and
found applications in many academic disciplines outside the traditional confines of physics. These include
sociology \cite{Galam}, economics \cite{Stefan}, complex networks  \cite{networks} as well as in more exotic areas \cite{football}.
Each of these disciplines involve cooperative phenomena emerging from the interactions between individual units.
Microscopic physical models  -- mostly of a rather simple nature -- help explain how the  
properties of such complex systems arise from the  properties of their individual parts.

Let ${b_g}_{\langle{i,j}\rangle}$ represent the strength of interaction $\langle{i,j}\rangle$ 
between the $i^{\rm{th}}$ and $j^{\rm{th}}$
individuals of group $g$. 
For a  group of size $N_g$,  the number of two-way communication links is $N_g(N_g-1)/2$. 
If all of these communication links are active,  Eq.(\ref{wrong}) should be replaced by 
\begin{equation}
 S_g = \sum_{i=1}^{N_g}{{a_g}_i} + \sum_{\langle{i,j}\rangle=1}^{N_g(N_g-1)/2}{{b_g}_{\langle{i,j}\rangle}} = N_g \bar{a}_g 
+ \frac{1}{2} N(N-1) \bar{b}_g,
\label{correct}
\end{equation}
where  $\bar{b}_g$ is the average strength of the intra-group interactions.

However, since two-way communication can only be  carried out effectively
between a limited number of individuals, one may further expect that, 
upon increasing the group size, a saturation or breakpoint point 
is eventually reached, 
beyond which the group fragments into sub-groups. We denote this breakpoint by  $N_c$.

There may also exist communication between such subgroups. 
Suppose the mean size of  subgroups in group $g$ is $\alpha_g N_c$,
and suppose the average inter-subgroup interaction strength for group $g$ is denoted by $\beta_g$. Then
the strength of the $g^{\rm{th}}$ group is 
\begin{equation}
 S_g =  N_g \bar{a}_g + \frac{1}{2} N_g(\alpha_g N_c-1) \bar{b}_g + \frac{\beta_g}{2}\frac{N_g}{\alpha_g N_c}
\left({ \frac{N_g}{\alpha_gN_c}-1}\right).
\label{correct2}
\end{equation}

In Eqs.(\ref{correct}) and(\ref{correct2}), the parameters $\bar{a}_g$, $\bar{b}_g$, $\alpha_g$ and $\beta_g$ represent
features of individual groups labelled by $g$. Averaging these values gives a representation of the
expected or average behaviour of the groups in the discipline to which group $g$ belongs.
We denote these averages as  $a$, $b$, $\alpha$ and $\beta$, respectively.  

Gathering terms of equal order, we therefore
 find that the average or expected strength $S$ of a group of size $N$ in a given discipline is
\begin{equation}
 S = \left\{ \begin{array}{ll}
             \left({ a - \frac{b}{2}}\right)N + \frac{b}{2} N^2 &  {\mbox{if $N \le N_c$}} \\
             \left({ a + \frac{b}{2}(\alpha N_c-1) - \frac{\beta}{2\alpha N_c}}\right)N + \frac{\beta}{2\alpha^2 N_c^2} N^2 &  {\mbox{if $N \ge N_c$}}.
             \end{array}
     \right.
\label{Ncccc}
\end{equation}
The $N$-dependency of the expected quality $s=S/N$ of research teams may therefore be considered as a measurement of the relative
importance of cooperation or collaboration within a given discipline. 
Beyond $N_c$, two-way communication between all team members is no longer the dominant driver
of research quality leading to a  milder dependency of $s$ on $N$ .
This phenomenon is akin to phase transitions in physics \cite{KeBe10}.
The crossover or breakpoint point $N_c$ may  be considered a demarkation 
point between ``small/medium'' and ``large'' groups 
in a given discipline and a measurement of the number of colleagues with whom a given
individual can collaborate in a meaningful sense.

In an attempt to capture the essence of this behaviour, and to measure $N_c$ for different disciplines,
a piecewise linear regression analysis is applied to data sets corresponding to different academic disciplines. 
We fit to the form 
\begin{equation}
 s = \left\{ \begin{array}{ll}
             a_1 + b_1 N &  {\mbox{if $N \le N_c$}} \\
             a_2 + b_2 N &  {\mbox{if $N \ge N_c$}},
             \end{array}
     \right.
\label{Nc}
\end{equation}
where $a_1$, $b_1$, $a_2$ and $b_2$ are related to the parameters appearing in (\ref{Ncccc}).
Note in particular that $b_2 \propto 1/N_c^2$ so that the slope to the right of the breakpoint should be small for
disciplines with large breakpoint values. Breakpoint values $N_c$ are 
also estimated through the fitting algorithm.

Having defined  {\emph{large}} teams as those whose size $N$ exceeds the breakpoint $N_c$, we next 
attempt to quantify the  meaning of the hitherto vague term ``critical mass'' \cite{Ha09}.  
This is  described as the value $N_k$ of $N$ beneath which research groups are vulnerable in the longer term.
We refer to teams of size $N < N_k$ as {\emph{small}} and groups with  $N_k < N <N_c$ as {\emph{medium}}.  
To determine $N_k$, one may ask, if funding is available to support new staff in a certain discipline, 
is more globally beneficial to allocate them to a small, medium or large team?

The addition of $M$ staff members to a research group described by (\ref{Nc}) may be expected to lead to an average increase 
in {\emph{absolute}} strength of 
\begin{equation}
 \frac{\Delta S}{M} = \frac{S(N+M)-S(N)}{M} = a_i+b_i(2N+M),
\label{b0}
\end{equation}
depending on whether $N<N_c$ (where $i=1$) or $N>N_c$ ($i=2$).
Some elementary algebra shows it is more beneficial if these new staff are allocated to the small/medium group if
$
 N > ({a_2 - a_1})/{2(b_1-b_2)} - M/2
$.
Taking the $M \rightarrow 0$ limit of this expression gives what may be considered the critical mass
of the research area, 
\begin{equation}
 N_k =  \frac{a_2 - a_1}{2(b_1-b_2)} = \frac{N_c}{2}
\,.
\label{Nk}
\end{equation}
(In a similar manner one can consider the loss of $M$ researchers from
a group of size $N$: in fact what we have done here is simply maximised the gradient of $S=sN$.)
We shall see that the slope $b_2$ to the right of the breakpoint $N_c$ is indeed small and
tends to vanish for large group size \cite{Ha09}. This being the case, from (\ref{b0}), the increase in total strength 
of a group by the addition of one more member is additive. The slope $b_1$ to the left of the breakpoint, however,
will be seen to be positive, so that the increase in strength  there is proportional to $N$.

Here we have considered the question of where to allocate $M$ new staff, should they become available to 
a particular research discipline. We may also ask the complementary question: if the total number of staff 
in a given area is fixed, what is the best strategy, on average, for transferring them between small/medium and large groups.
It turns out that incremental transfer of staff from a large group to a  medium one (one which already exceeds critical mass)
increases the overall strength of the discipline \cite{KeBe10}. 
A sub-critical group must, however, achieve critical mass before such a move is globally beneficial.

To summarize, based upon the notion that two-way collaborative links are the main drivers of quality,
we have developed a model which classifies research groups as  small, medium and large. 
If external funding becomes available, it is most beneficial to support groups which are medium in size.
Likewise, the incremental transfer of staff from large groups to medium ones promotes overall research quality \cite{KeBe10}.
On the other hand, small groups must strive to achieve critical mass to avoid extinction.
Our approach involves piecewise linear models since these are easily 
interpreted as representing the effects of collaboration and of pooling of resources. 
We consider other models (such as higher-degree polynomial fits) less appropriate as they are not so easily interpreted.
Similarly, although the opposite causal direction (quality leads to increase in group size) 
no doubt plays an important role in the evolution of research teams, this is less straightforward
to interpret in a linear manner. (Indeed on the basis of the empirics we shortly present, 
we shall see that this is not the  dominent driver of group evolution.)
 With these caveats in mind, we proceed to report on the analysis of various subject areas as evaluated in the
UK's RAE.

\section{Analysis of various subject areas}

We begin the analysis with the subject area with which we are most familiar, namely applied mathematics, 
which includes some theoretical physics groups (Fig.~1(b)). 
For applied mathematics, the smallest group consisted of one individual and the largest had $80.3$, 
with the mean group size being $18.9$. (Fractional staff numbers are associated with part-time staff.)
As stated, the first observation  from the figure is that the quality $s$ indeed tends to increase with group size $N$.
The solid curve in Fig.~1(b) is a piecewise linear regression fit to the data and the dashed curves represent the 
resulting 95\% confidence belt for the normally distributed data.
The fitted breakpoint $N_c=12.5 \pm 1.8$ splits the 45 research teams into 16 small/medium groups 
and 29 large ones.
The value for the critical mass is calculated from (\ref{Nk}) to be $N_k = 6.2 \pm 0.9$ which, based on experience, we consider to 
be a reasonable value for this subject area.

\begin{figure}[t]
\begin{center}
\hspace{-1cm}
\includegraphics[width=0.45\columnwidth, angle=0]{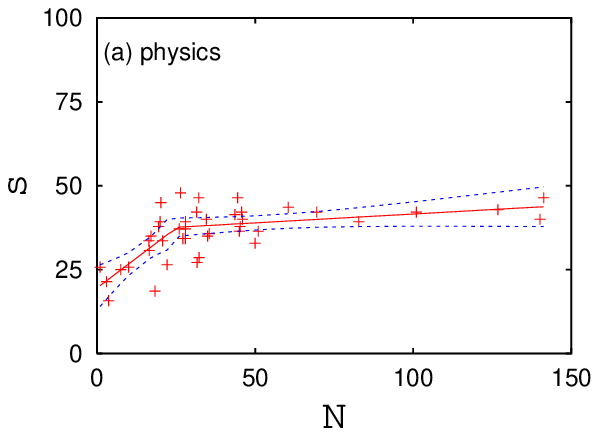}
\includegraphics[width=0.45\columnwidth, angle=0]{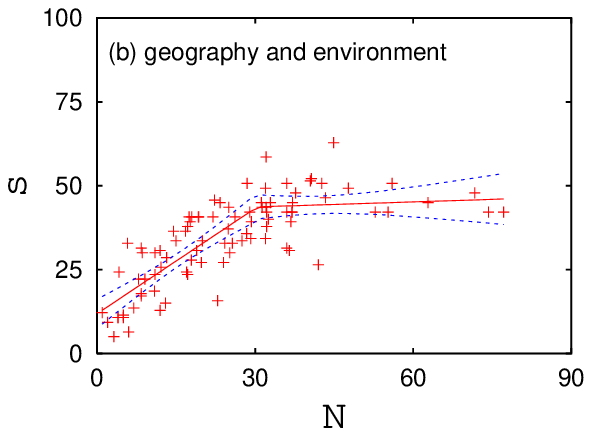}
\\
\hspace{-1cm}
\includegraphics[width=0.45\columnwidth, angle=0]{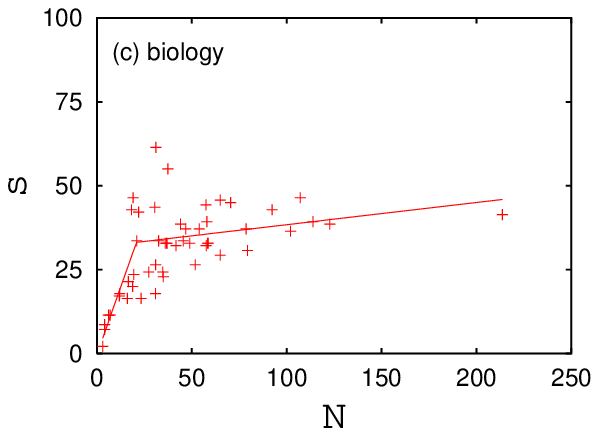}
\includegraphics[width=0.45\columnwidth, angle=0]{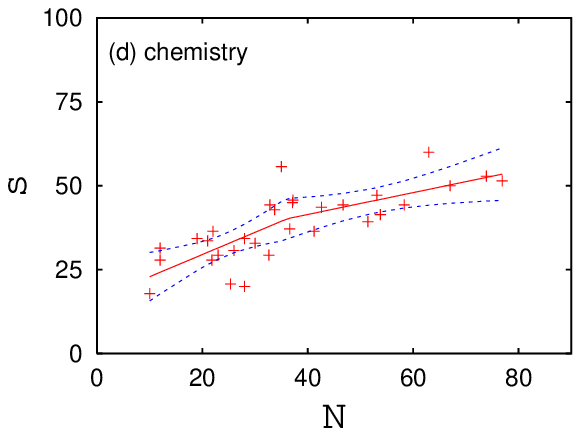}
\caption{Research quality $s$ as a function of group quantity $N$ for 
(a) Physics,
(b) Geography,  Earth and environmental  sciences,
(c) Biology, and for 
(d) Chemistry.
}
\end{center}
\end{figure}
Closely related to applied mathematics is the physics unit of assessment, which had 42 RAE submissions. 
While some theoretical physics teams were submitted under this banner, the bulk of the unit of assessment is 
composed of experimentalists. 
The analysis is presented in Fig.~2(a) and the statistics summarizing this and other fits are listed 
in Table~1 in which $n$ represents the sample size (number of groups) and $R^2$ is the coefficient of determination.
For physics, the critical mass is estimated to be $N_k = 12.7 \pm 2.4$, about twice that  for applied 
mathematics and closer to those  for geography, Earth and environmental sciences
and for  biology. 
These data for these natural sciences are plotted in Fig.~2(b) and 2~(c), respectively.
The main results of these analyses are also listed in Table~1. 

In Table~1, various statistical indicators are listed. 
The $P_{\rm{m}}$ are the $P$-values for the null hypothesis that there is no underlying correlation
between $s$ and $N$. Also, $P_{b_1-b_2}$ are the $P$-values for the hypothesis that 
the slopes coincide on either side of the breakpoint.
Small values of these indicators (below $0.05$, say) indicate that we may reject these null hypotheses.
The large value of $P_{b_1-b_2}$ for chemistry (Fig.~2(d)) 
suggests that the null hypothesis of coinciding slopes cannot be rejected, 
resulting in the large error bar reported for the critical mass in Table~1.
Indeed, a single-line fit to all chemistry data yields a $P$-value for the 
model of less than $0.001$ and $R^2=59.6$. 
In this case the size of the smallest group submitted to RAE was $N=10$, 
which is larger than for the other subjects considered
(the minimum group size in physics was $N=1$, and for biology it was $N=3$)
and indicates that chemistry in the UK is dominated by medium/large groups, most small ones having already petered out.

\begin{figure}[t!]
\begin{center}
\hspace{-1cm}
\includegraphics[width=0.45\columnwidth, angle=0]{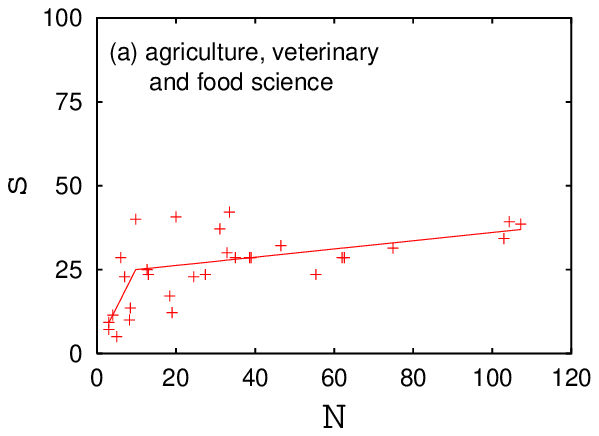}
\includegraphics[width=0.45\columnwidth, angle=0]{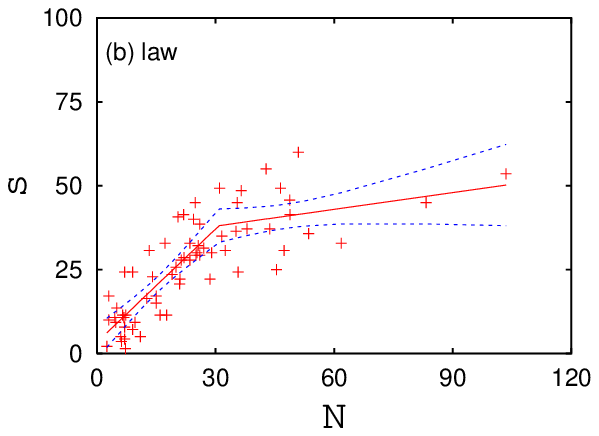}
~\\
\hspace{-1cm}
\includegraphics[width=0.45\columnwidth, angle=0]{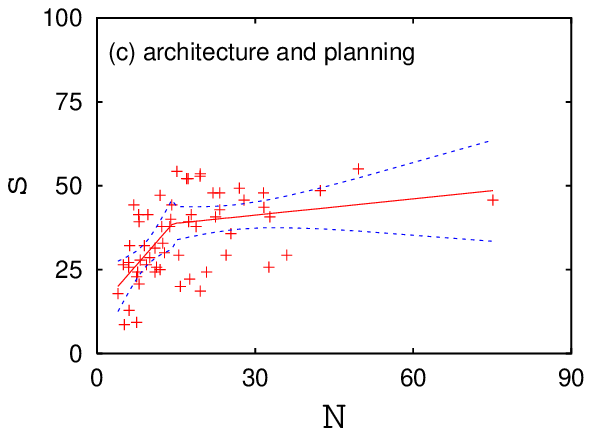}
\includegraphics[width=0.45\columnwidth, angle=0]{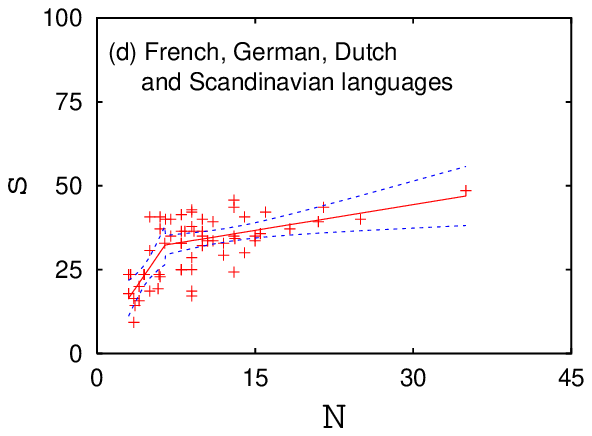}
~\\
\hspace{-1cm}
\includegraphics[width=0.45\columnwidth, angle=0]{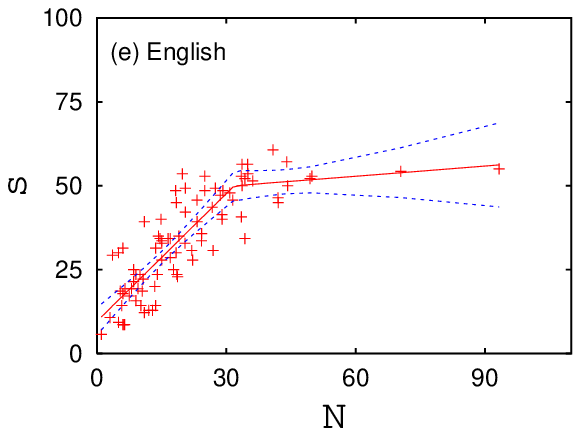}
\includegraphics[width=0.45\columnwidth, angle=0]{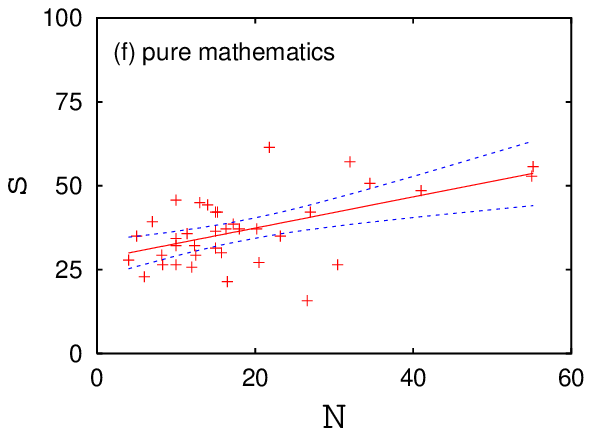}
\caption{Quality $s$ as a function of group size $N$ for various academic disciplines.
(a) Agriculture, veterinary and food science.
(b) Law.
(c) Architecture and town and country planning.
(d) French, German, Dutch and Scandinavian languages.
(e) English language and literature.
(f) Pure mathematics, where no breakpoint was detected.
}
\end{center}
\end{figure}
Besides the natural sciences, we have analysed work patterns in many other academic disciplines
and listed the results in Table~1. Some of these are depicted in Fig.~3.
All the data listed support our models, except for pure mathematics which we will address shortly.
In the table, we indicate the cases which do not entirely satisfy tests of normality. 
In particular, the Kolmogorov-Smirnov test for normality fails for law, which however passes the Anderson-Darling test.
The following areas fail both normality tests: biology; agriculture, veterinary and food science;
politics and international studies; history; art and design. 
In these cases the $P$ values and confidence intervals can only be considered approximate and we omit confidence intervals from 
Figs.~2(c) and~3(a). 

\begin{table}[!t]
\caption{The results of the analysis of the RAE for a variety of academic disciplines.
Here $n$ represents the number of research groups in each area, the critical mass for which is estimated to be $N_k$.
The coefficients of determination are $R^2$.
The $P$-values are $P_{m}$ for the absence of correlation between $s$ and $N$ for all data,
$P_{b_1-b_2}$ for coinciding slopes on either side of the breakpoint $N_c$, and 
 $P_{b_2}$ for the absence of correlation for the large groups.
The symbol $\dagger$ indicates that law fails the Kolmogorov-Smirnov test for normality, 
but passes the Anderson-Darling test. 
The symbols $\ddagger$ indicate failures of both normality tests.  
The symbol $\ast$ indicates that pure mathematics is best fitted by a single line.}
\begin{center}
\begin{tabular}{|l|r|r|r|r|r|r|} \hline \hline
Subject                                  &$n$  & $N_k=N_c/2$    & $R^2$    & $P_{\rm{m}}$ & $P_{b_1-b_2}$ & $P_{b_2}$\\
                                         &     &                &          &              &               &    \\
\hline
Applied mathematics                      & 45  & $6.2 \pm 0.9$  & $74.3$   & $<0.001$  &  $<0.001$  &  $ 0.001 $     \\
Physics                                  & 42  &$12.7 \pm 2.4$  & $53.0$   & $<0.001$  &  $ 0.003$  &  $ 0.098 $        \\
Geography, Earth \& environment          & 90  & $15.2 \pm 1.4$ & $65.9$   & $<0.001$  &  $<0.001$  &  $ 0.627 $        \\
Biology$^{\ddagger}$                     & 51  & $10.4 \pm 1.6$ & $53.7$   & $<0.001$  &  $<0.001$  &  $ 0.096 $        \\
Chemistry                                & 31  & $18.1 \pm 6.4$ & $62.1$   & $<0.001$  &  $ 0.206$  &  $ 0.026 $        \\
Agriculture, veterinary, etc$^{\ddagger}$& 30  & $ 4.9 \pm 1.4$ & $52.3$   & $<0.001$  &  $ 0.115$  &  $ 0.045 $        \\
Law$^{\dagger}$                          & 67  & $15.4 \pm 1.9$ & $70.8$   & $<0.001$  &  $<0.001$  &  $ 0.113 $        \\
Architecture \& planning                 & 59  &$ 7.1 \pm 1.4$  & $33.4$   & $<0.001$  &  $ 0.014$  &  $ 0.261 $        \\
French, German, Dutch                    & 62  & $3.2 \pm 0.4$  & $49.8$   & $<0.001$  &  $ 0.004$  &  $ 0.008 $        \\
                      \& Scandanavian    &     &                &          &           &            &             \\
English language and literature          & 87  &$15.9 \pm 1.4$  & $73.6$   & $<0.001$  &  $<0.001$  &  $ 0.407 $        \\
Pure mathematics$^\ast$                  & 37  & $ \le 2 $      & $29.1$   & $<0.001$  &            &             \\
Medical sciences                         & 82  &$20.4 \pm 4.0$  & $27.5$   & $<0.001$  &  $ 0.006$  &  $ 0.118 $        \\
Nursing, midwifery etc.                  & 103 & $ 9.2 \pm 2.2$ & $19.7$   & $<0.001$  &  $ 0.017$  &  $ 0.364 $        \\
Computer science 1                       & 81  & $ 24.6 \pm 5.0$& $45.0$   & $<0.001$  &  $ 0.007$  &  $ 0.954 $        \\
Computer science 2                       & 81  & $ 16.3 \pm 4.3$& $43.3$   & $<0.001$  &  $ 0.014$  &  $ 0.065 $        \\
Computer science 3                       & 81  & $  5.6 \pm 2.4$& $40.9$   & $<0.001$  &  $ 0.252$  &  $<0.001 $        \\
Archaelogy 1                             & 26  & $12.7 \pm 1.6$ & $74.9$   & $<0.001$  &  $<0.001$  &  $ 0.816 $        \\
Archaelogy 2                             & 26  & $ 8.5 \pm 1.2$ & $74.7$   & $<0.001$  &  $<0.001$  &  $ 0.154 $        \\
Economics \& econometrics                & 35  & $ 5.3 \pm 1.4$ & $59.1$   & $<0.001$  &  $ 0.091$  &  $<0.001 $        \\
Business \& management                   & 90  &$23.8 \pm 3.8$  & $60.4$   & $<0.001$  &  $<0.001$  &  $ 0.042 $        \\
Politics \& international 
          studies$^{\ddagger}$           & 59  &$12.5 \pm 2.1$  & $53.8$   & $<0.001$  &  $ 0.001$  &  $ 0.115 $        \\
Sociology                                & 39  &$ 7.0 \pm 1.6$  & $50.7$   & $<0.001$  &  $ 0.086$  &  $ 0.010 $        \\
Education                                & 81  &$14.5 \pm 2.2$  & $55.7$   & $<0.001$  &  $<0.001$  &  $ 0.336 $        \\
History$^{\ddagger}$                     & 83  &$12.4 \pm 2.3$  & $49.7$   & $<0.001$  &  $<0.001$  &  $ 0.054 $        \\
Philosophy \& theology                   & 80  &$ 9.5 \pm 1.5$  & $49.7$   & $<0.001$  &  $<0.001$  &  $ 0.574 $        \\
Art \& design$^{\ddagger}$               & 71  &$12.5 \pm 3.7$  & $19.6$   & $0.002$   &  $ 0.019$  &  $ 0.616 $        \\
History of art, performing arts,         &172  &$ 4.5 \pm 0.8$  & $27.5$   & $<0.001$  &  $ 0.011$  &  $ 0.006 $        \\
communication studies and music          &     &                &          &           &            &                   \\
\hline
\hline \hline
\end{tabular}
\end{center}
\end{table}

The results for pure mathematics are presented in Fig.~3(f) and merit special comment. 
In pure mathematics no breakpoint was detected and the data is best fitted by a single line $s=a+bN$. 
The relatively small slope and large intercept of this fit are similar to the corresponding values 
for {\emph{large}} groups in applied mathematics. This suggests that the set of the pure mathematics 
data may be interpreted as also corresponding to large groups. 
In this case $N_c$ may be interpreted as being less than or equal to the size of the
smallest group submitted to the RAE, which was $4$, so that the critical mass $N_k$ is $2$, or less. 
This indicates that local cooperation is less significant in pure mathematics, where the work pattern
is more individualised. This is consistent with experience: in pure mathematics publications  tend to be authored
by one or two individials, rather than by larger collaborations.

In computer science, three competing candidates for the breakpoint were found, each with similar coefficients of determination. 
These are  
$N_c=49.1 \pm 10.0$ with $R^2=45.0$, 
$N_c=32.5 \pm  8.5$ with $R^2=43.3$, and
$N_c=11.3 \pm  4.7$ with $R^2=40.9$. 
We interpret these as signaling that this discipline is in fact an amalgam of 
several sub-disciplines, each with their own work patterns and their own critical masses.
These are listed in the table as computer science 1, 2, and 3, respectively.
Similarly, for archaeology, which with 26 data points is the smallest 
data set for which we present results, 
besides the peak in $R^2$ corresponding a breakpoint at $N_c=25.4 \pm 3.2$ where $R^2=   74.9$, 
there is also a local maximum  at $N_c= 17.0 \pm 2.4$  where $R^2=   74.7$.
No other  discipline displays this feature of multiple breakpoints, 
which we interpret as signaling that these other fields are quite homogeneous in their work patterns.

In Table~1, $P_{b_2}$ are the $P$-values for the hypothesis that the large-group data are uncorrelated.
From the model developed in Sec.~2, one expects this saturation to be  triggered for sufficiently large values of $N_c$.
Indeed, this is the case for 16 subject areas to the right of the breakpoint, with $N_c > 14$.
On the other hand, the data for 8 subject areas do not appear to completely flatten to the right of $N_c$. 
Six of these have a relatively small value of the breakpoint, and, according to Sec.~2,
 we interpret the continued rise of $s(N)$ 
to the right of $N_c$ as being due to inter-subgroup cooperation \cite{KeBe10}.
(The cases of chemistry and business/management buck this trend in that the data rises to the right
despite having a comparably large breakpoint.)
Although we have insufficient large-$N$ data to test, one may expect that these cases will also ultimately saturate 
for sufficiently large $N$ \cite{KeBe10}.

The English funding formula was adjusted in 2010 in such a way that 
4* research receives a greater proportion of funds. The corresponding formula
is $ s = (9p_{4*} + 3p_{3*} + p_{2*})/9$. We have tested robustness of our analysis by 
checking that this change over the 2009 formula (Eq.(\ref{seven})) does not alter our conclusions 
within the quoted errors.  
For example, for applied mathematics, $N_k$ changes from $6.2 \pm 0.9$ with $R^2= 74.3$ 
to $N_k = 6.3 \pm 1.0$ with $R^2 = 73.5$. Similarly, the values for physics
change from $N_k = 12.7 \pm 2.4$ with $R^2=53.0$ to  $N_k = 12.6 \pm 2.3$ with $R^2=54.4$.
Another test of robustness in the applied mathematics case is to remove the two
 points with largest $N$ and $s$ values (In Fig.~2(b)) from the analysis. 
The resulting values are $N_k = 6.1 \pm 1.0$, with $R^2=71.5$, while the estimates
for $a_1,\dots a_2$ remain stable. 
Thus even the biggest and best groups follow the same patterns as the other large groups.

\begin{figure}[t]
\begin{center}
\hspace{-1cm}
\includegraphics[width=0.45\columnwidth, angle=0]{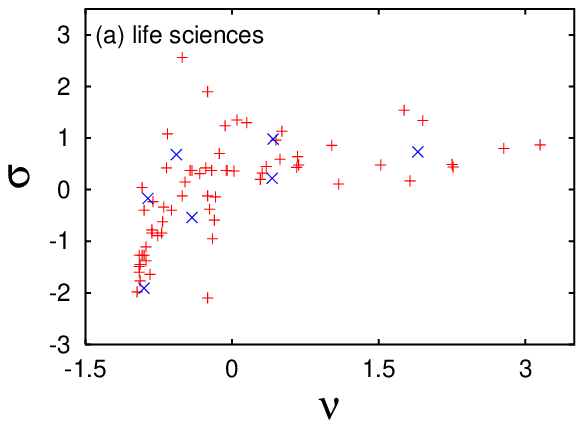}
\includegraphics[width=0.45\columnwidth, angle=0]{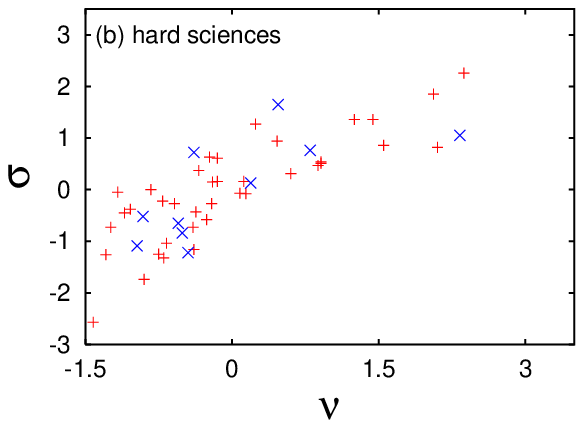}
\caption{ 
Standardised quality plotted against standardised sizes  for  the UK's RAE and France's AERES
for (a) the life sciences and (b) the hard sciences. 
Here  $\sigma  = (s - {\bar{s}})/\sigma_s$   $\nu = (N - {\bar{N}})/\sigma_N$.
The integrated British data are denoted by the symbols  $+$ (red online) and the French data  
are indicated by  $\times$ (blue online). }
\end{center}
\end{figure}
To illustrate our belief that the scenario described above is not only a UK phenomenon, we compare to the 
French AERES system for life sciences and hard sciences. 
As stated, the French data refers to results on a broader institutional scale and fewer data are available.
Therefore signals corresponding to smaller individual research groups are swamped and no breakpoint is 
expected to be  measureable. 
To perform the comparison, we apply the same formula as for the English system and in Fig.~4, we
plot $\sigma = (s-\bar{s})/\sigma_s$ against  $\nu =(N-\bar{N})/\sigma_N$ for both cases,
where $\bar{N}$ and $\bar{s}$ are the mean $N$ and $s$ values, respectively, for the integrated data
and $\sigma_N$ and $\sigma_s$ are their standard deviations. 
In the case of the life sciences (Fig.~4(a)), we check that the integrated British data, obtained by merging 
the biological sciences, pre-clinical and human biology with agriculture, veterinary and food sciences
yield results which overlap with their French counterparts, which they evidently do. 
The corresponding plot for the hard sciences (in Fig.~4(b)), in which for the British data we have integrated  
pure and applied mathematics, statistics, physics, Earth sciences, chemistry, computer science and informatics 
also overlap convincingly.

\section{Discussion}

\begin{figure}[t]
\begin{center}
\includegraphics[width=0.45\columnwidth, angle=0]{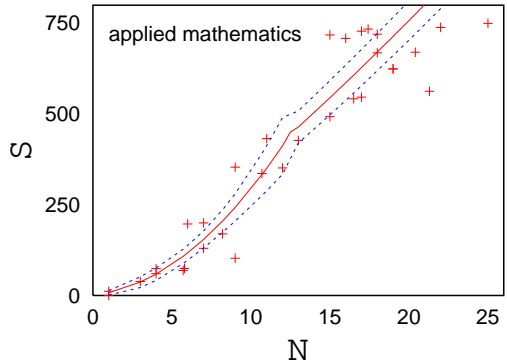}
\caption{The absolute strenght $S=sN$ for applied mathematics near the breakpoint $N_c=12.5$, where the gradient is maximised. 
}
\end{center}
\end{figure}
To reiterate our main point concerning the correlation between the quality and group sizes, we plot in 
Fig.~5 the absolute strength values  $S$ for applied mathematics as a function of $N$ in a region around the
breakpoint, to illustrate the maximal gradient there. 
Without breakpoint, the steepest gradient condition would lead to supporting only the biggest among the large groups
so that only one group remains at the end.
The resulting instability is described in Ref.~\cite{Ha09} in the case of the former Soviet Union,
where the scenarios oscillated between this and the alternative one.

At the end of Section~2, we mentioned that the opposite causal mechanism of increasing quality 
driving increasing group size, may be dismissed as the dominant mechanism on the basis of empirical evidence.
Such a model may be considered at the microscopic level by preferential attachment of quality to quality (a ``success-breeds-success'' mechanism). In such a system, a high quality group may gain more funding and find
it easier to attract more high quality researchers, and hence may grow while maintaining quality.
The converse may hold for a lower quality group. 
However,   if this were the primary mechanism, one would expect a monotonic increase of $N$ with $s$ to plateau
only when the maximum possible quality  $s=1$ is achieved, and the absence of a phase transition prior to this point. 
Since none of the disciplines analysed exhibit such behaviour, we may dissmiss it  as the primary causal
link between quality and quantity, at least for the subject areas presented.
However, undoubtedly such a feedback mechanism may be at work at a sub-dominant level.
 
\begin{figure}[t]
\begin{center}
\includegraphics[width=0.45\columnwidth, angle=0]{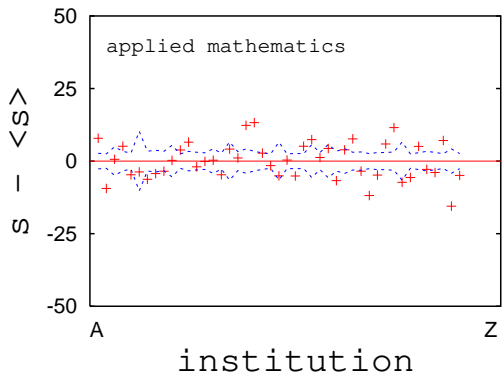}
\caption{The renormalised quality  $s-\langle{s}\rangle$ for applied mathematics, where $\langle{s}\rangle$ is the 
quality predicted by the model. The tighter bunching of the data compared to the raw plot of Fig.~1(b) (the two figures are to the same
scale)
illustrates the validity of the model and that the averaged individualised performances are better than hitherto realised.
}
\end{center}
\end{figure}

Finally, we discuss an important consequence of our analysis regarding perceptions of individual research calibre
resulting from exercises such as the RAE and AERES.
As discussed in Sec.~2, 
Fig.~1(a) invites a comparison of group performance, from which it is tempting to deduce  
information on the average calibre of individuals forming those groups.
The correspondingly naive interpretation associated with Eq.(\ref{wrong})
leads to the {\emph{erroneous}} conclusion that $\bar{a}_g = s_g$,
i.e., that 
the average strength of individuals in group $g$ is given by the measured quality of that group.

We have seen that this argumentation is {\emph{incorrect}} because it fails to take intra-group interactions into account --
i.e., it fails to take account of the complex nature of research groups, the importance of which we have demonstrated.

From Eq.(\ref{correct}), 
the average strength of {\emph{individuals}} in group $g$ is in fact given by 
\begin{equation}
 \bar{a}_g = s_g - (N-1) \bar{b}_g.
\label{right}
\end{equation}
The group-size-dependent second term swamps the relationship between group quality and 
average individual strength, illustrating how dangerous (and wrong) it is to draw conclusions
on individual calibre from group measurements.
A similar conclusion may be drawn for large groups from Eq.(\ref{correct2}).

A more accurate represention of the average quality of {\emph{individuals}} in given research groups
is illustrated in Fig.~6, where, for applied mathematics, we plot a  renormalised quality  $s-\langle{s}\rangle$ 
against $N$. Here $\langle{s}\rangle$ is the quality  expected from the model (\ref{Nc}). 
Subtracting it from the measured quality $s$ has the effect of removing the average strengths of two-way collaborations.
The deviations of data points from the line therefore represent their comparisons to the {\emph{local}} average (i.e. to
their positions predicted by the model).
The plot is on the same scale as that for the raw data in Fig.~1(a), where the line represents the 
{\emph{global}} average to facilitate comparison between the incorrect naive approach and 
the one based on Eq.(\ref{right}).
The standard deviation for Fig.~6 is $6.4$, about half the value of $12.6$ for 
Fig.~1(a). Thus the renormalised data are distributed closer to the local  expectation values than the 
raw data are to the global average. This illustrates that the averaged individualised RAE performances of small and medium
research groups are even better than hitherto realised and that small and medium groups indeed host pockets of excellence.

\section{Conclusions}

By taking seriously the notion that research groups are complex interacting systems,
our study explains why  the average quality of bigger groups appears to exceed that of smaller groups.
It also shows that it is unwise to judge groups - or individuals within groups - solely on the basis of  quality profiles,
 precisely because of this strong size dependency. Because of the overwelming importance of two-way communication links, 
small and medium sized groups should not be expected to yield
the same quality profiles as large ones, and to compare small/medium groups to the average quality over 
all research groups in a given discipline would be misleading. 
An analysis of the type presented here may therefore assist 
in the determination of which groups are, to use a boxing analogy, punching above or below their weight within a 
given research arena.
This type of analysis should be taken into account by decision makers when 
comparing research groups and when formulating strategy.
Indeed we have shown that to optimize overall research performance in a given discipline,
medium sized groups should be promoted while small ones must endeavour to attain critical mass.

Furthermore, we have compared the French and UK research evaluation systems and found them to be
consistent, although the RAE gives information at a finer level that does the AERES. 

Finally, we have quantified the hitherto intuitive notion of {\emph{critical mass}\/} in research and determined 
their values for a variety of different academic disciplines.


\bigskip
%

\vspace{1cm}
\noindent
{\bf{Acknowledgements}} 
We are grateful to Neville Hunt for inspiring discussions as well as for help with the statistical analyses. 
We also thank Arnaldo Donoso, Christian von Ferber, Housh Mashhoudy and Andrew Snowdon for comments and discussions,
as well as Claude Lecomte, Scientific Delegate  at the AERES, for discussions on the work of that agency.

\end{document}